\documentclass[aps,prl,reprint,letterpaper,amsmath,amssymb,longbibliography,superscriptaddress]{revtex4-2}
\usepackage{amsmath}
\usepackage{graphicx}

\begin{document}
\title{Temperature-dependent photoluminescence dynamics of CsPbBr$_3$ and CsPb(Cl,Br)$_3$ perovskite nanocrystals in a glass matrix}
\author{E.~V.~Kulebyakina}
\affiliation{P.N. Lebedev Physical Institute of the Russian Academy of Sciences, 119991 Moscow, Russia}
\author{M.~L.~Skorikov}
\affiliation{P.N. Lebedev Physical Institute of the Russian Academy of Sciences, 119991 Moscow, Russia}
\author{E.~V.~Kolobkova}
\affiliation{St. Petersburg State Institute of Technology (Technical University), 190013 St. Petersburg, Russia} 
\affiliation{ITMO University, 199034 St. Petersburg, Russia}
\author{M.~S.~Kuznetsova}
\affiliation{Spin Optics Laboratory, St. Petersburg State University, 198504 St. Petersburg, Russia}
\author{M.~N.~Bataev}
\affiliation{Spin Optics Laboratory, St. Petersburg State University, 198504 St. Petersburg, Russia}
\author{D.~R.~Yakovlev}
\affiliation{P.N. Lebedev Physical Institute of the Russian Academy of Sciences, 119991 Moscow, Russia}
\affiliation{Experimentelle Physik 2, Technische Universit\"{a}t Dortmund, 44221 Dortmund, Germany}
\affiliation{Ioffe Institute, Russian Academy of Sciences, 194021 St. Petersburg, Russia}
\author{V.~V.~Belykh}
\email[]{E-mail: vasilii.belykh@tu-dortmund.de}
\affiliation{Experimentelle Physik 2, Technische Universit\"{a}t Dortmund, 44221 Dortmund, Germany}

\begin{abstract}
Lead halide perovskite nanocrystals (NCs) in a glass matrix combine excellent optical properties and stability against environment. The spectral and temporal characteristics of photoluminescence from CsPbBr$_3$ and CsPb(Cl,Br)$_3$ nanocrystals (NCs) in a fluorophosphate glass matrix are measured in a temperature range from 6 to 270~K in order to reveal factors that determine their quantum yield and recombination dynamics. At low temperatures, the recombination dynamics is characterized by three decay components with time scales on the order of 1~ns, 10~ns, and 1~$\mu$s. The relative contributions of the corresponding processes and their characteristic times are strongly temperature dependent. The emission intensity decreases with growing temperature. This effect is stronger in smaller NCs, which highlights the role of surface states. These experimental results are discussed on the basis of a model taking into account the NC energy structure and the presence of electron and hole surface trap states. The photoluminescence dynamics at low temperatures is dominated by charge-carrier radiative recombination and relaxation to shallow traps. At temperatures exceeding 100 K, the dynamics is affected by carrier activation to the excited states.
\end{abstract}
\maketitle


\section{Introduction}
\begin{figure*}
\includegraphics[width=1.0\linewidth]{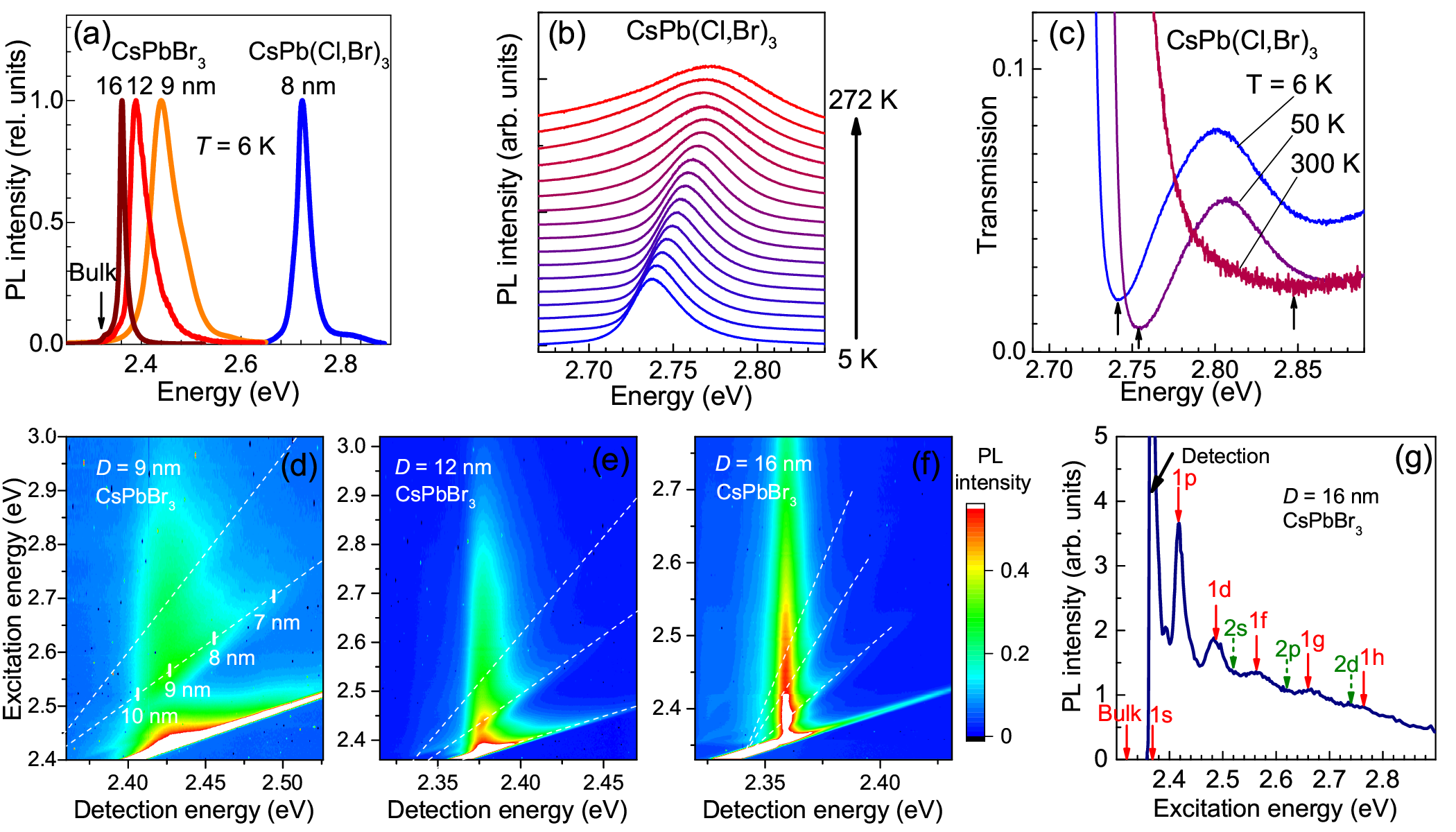}%
\caption{(a) PL spectra of CsPbBr$_3$ and CsPb(Cl,Br)$_3$ NCs with different sizes in a fluorophosphate glass matrix at $T = 6$~K. Arrow shows the exciton energy
in bulk CsPbBr$_3$, equal to 2.32 eV. (b) PL and (c) transmission spectra of CsPb(Cl,Br)$_3$ NCs at different temperatures. PL spectra are normalized to the respective peak values and shifted vertically for clarity. Arrows indicate transmission minima corresponding to the exciton resonance. (d)--(f) PL excitation spectrum maps for CsPbBr$_3$ NCs of different sizes showing the PL intensity as a function of the detection energy and excitation energy. Dashed lines are guides for the eye. (g) PL excitation spectrum at the detection energy of 2.37~eV for CsPbBr$_3$ NCs with size of 16~nm. Arrows show positions of energy levels calculated for spherical NC (see Modeling section).} 
\label{fig:PLTr}
\end{figure*}

Lead halide perovskite semiconductors are known for more than a century \cite{Wells1893,Moller1958}. However, only recently they emerged as a promising platform for photovoltaic applications \cite{Green2014},  which stimulated research in different directions revealing many advantages of perovskites. Among them are defect tolerance \cite{Huang2017}, high quantum yield \cite{Sutter-Fella2016}, high efficiency of spin orientation by light \cite{Belykh2019,Odenthal2017,Grigoryev2021,KirsteinLead2022,KirsteinML2023}, ease of synthesis, and the possibility to form nanocrystals (NCs) \cite{Vinattieri2021,Vardeny2022}. In particular, perovskite NCs can be very promising for light-emitting devices \cite{Protesescu2015}. They have relatively narrow emission spectrum with central wavelength controlled by the anion composition and NC size. However, colloidal lead halide perovskite NCs have rather poor stability when exposed to air, humidity, elevated temperatures, or intense light. A promising approach to solve this problem is to synthesize all-inorganic lead halide perovskite NCs embedded in glass \cite{Liu2018,Liu2018a,Li2017,Ai2016,Ai2017,Ye2019}. Especially suitable is the fluorophosphate glass matrix \cite{Kolobkova2021}, which offers high chemical resistance to harmful environmental conditions and the ability to introduce high concentrations of halides ensuring also high quantum yield of NCs.

For the further development of perovskite-based light-emitting devices it is crucial to get physical insight into processes that limit the quantum yield and radiative recombination rate in perovskite NCs in a glass matrix and to clarify how these parameters depend on the NC size and composition. To this end, it is straightforward to study spectra and dynamics of photoluminescence (PL) from NCs as a function of temperature. The temperature-dependent properties of the continuous-wave PL spectra of CsPbBr$_3$ NCs in a glass matrix at $T=40-240$~K were investigated in Ref.~\cite{Ai2017}, where, in particular, the activation-like decrease in the PL  intensity and the broadening of the PL line with temperature were observed. 

In this paper, we present systematic studies of the PL spectra and recombination dynamics down to liquid-helium temperatures for CsPbBr$_3$ and CsPb(Cl,Br)$_3$ NCs of different sizes embedded into a fluorophosphate glass matrix. On the basis of a model involving surface traps, we discuss factors responsible for PL decrease with temperature increase and for the observed complex PL dynamics. We find that an increase in the NC size leads to the enhancement of their spectral homogeneity and makes the PL intensity less sensitive to temperature. At the same time, with the introduction of Cl, PL temperature quenching becomes more pronounced.

\section{Experimental details}
\textbf{Samples} of fluorophosphate glass with composition of 40P$_2$O$_5$--35BaO--5NaF 10AlF$_3$--10Ga$_3$O$_3$ (mol.~\%) doped with NaCl (Ba$_2$Cl), Cs$_2$O, PbF$_2$ and BaBr$_2$ were synthesized using the melt-quench technique. Glass synthesis was performed in a closed glassy carbon crucible at a temperature $T = 1000 ^\circ$C. About 50~g of the batch was melted in a crucible for 30~min, then the glass melt was cast on a glassy carbon plate and pressed to form a plate with a thickness of 2~mm. CsPbBr$_3$ or CsPb(Cl,Br)$_3$ perovskite NCs were precipitated by glass self-crystallization during melt-quenching and additional heat treatment at $400^\circ$C. The size of NCs in the samples varied from 8 to 16 nm depending on the annealing time. The composition of the sample with chlorine was evaluated using the X-ray diffraction method to be CsPb(Cl$_{0.5}$Br$_{0.5}$)$_3$  \cite{Kirstein2022}.

For \textbf{PL} measurements, the samples were placed in a helium vapor-flow cryostat to achieve temperatures from 5 to 270~K.  The steady-state PL spectra were recorded with a resolution of 0.9~meV using a grating spectrometer equipped with a liquid-nitrogen-cooled charge-coupled-device (CCD) matrix detector. The sample was excited using a CW semiconductor laser with a wavelength of 405~nm and a power of 2--5~$\mu$W. The size of the laser spot on the sample was about 200~$\mu$m.

An incandescent tungsten ribbon lamp was used to measure the optical \textbf{transmission spectra}. The investigated sample was thinned to 43~$\mu$m. The spectra were recorded using the same grating spectrometer with a liquid-nitrogen-cooled CCD detector.

For the investigation of the \textbf{PL excitation spectra}, the samples were placed in a helium closed-cycle cryostat and cooled to a temperature of $12$~K. The PL was recorded using a  spectrometer equipped with a liquid-nitrogen-cooled CCD camera. Optical excitation was carried out using light from an incandescent lamp transmitted through a pre-monochromator and focused onto the sample in a spot with a diameter of 2 mm. The excitation power was about 0.4 $\mu$W. The experimental data were normalized to the power of the incident light at a given wavelength. The spectral resolution of the PL excitation spectra was determined by the FWHM of the light band transmitted through the pre-monochromator, which was about 7 meV (1.5 nm).

The \textbf{PL dynamics} was measured with a Hamamatsu C5680 streak camera coupled to a grating spectrograph. The spectral resolution was typically about 4~meV, and the time resolution was about 0.01 of the measurement time range. In these measurements, the sample was excited at a wavelength of 400~nm by the second harmonic of 2.5-ps pulses from a Mira-900D mode-locked Ti:sapphire laser with a pulse repetition rate of 76~MHz, which could be lowered by a factor of 16--8192 using a pulse picker for time-resolved measurements in time ranges up to 100~$\mu\text{s}$. The diameter of the laser spot on the sample was about $0.5 \times 0.3$~mm and the pulse energy was about 0.5 pJ (corresponding to an average power of about 2~$\mu\text{W}$ for a pulse repetition rate of 4.75~MHz). We checked that we work in the linear regime, i.e. the shape of the PL dynamics is independent of the excitation power.

\section{Photoluminescence under continuous-wave excitation}

\begin{figure*}
\begin{center}
\includegraphics[width=0.7\linewidth]{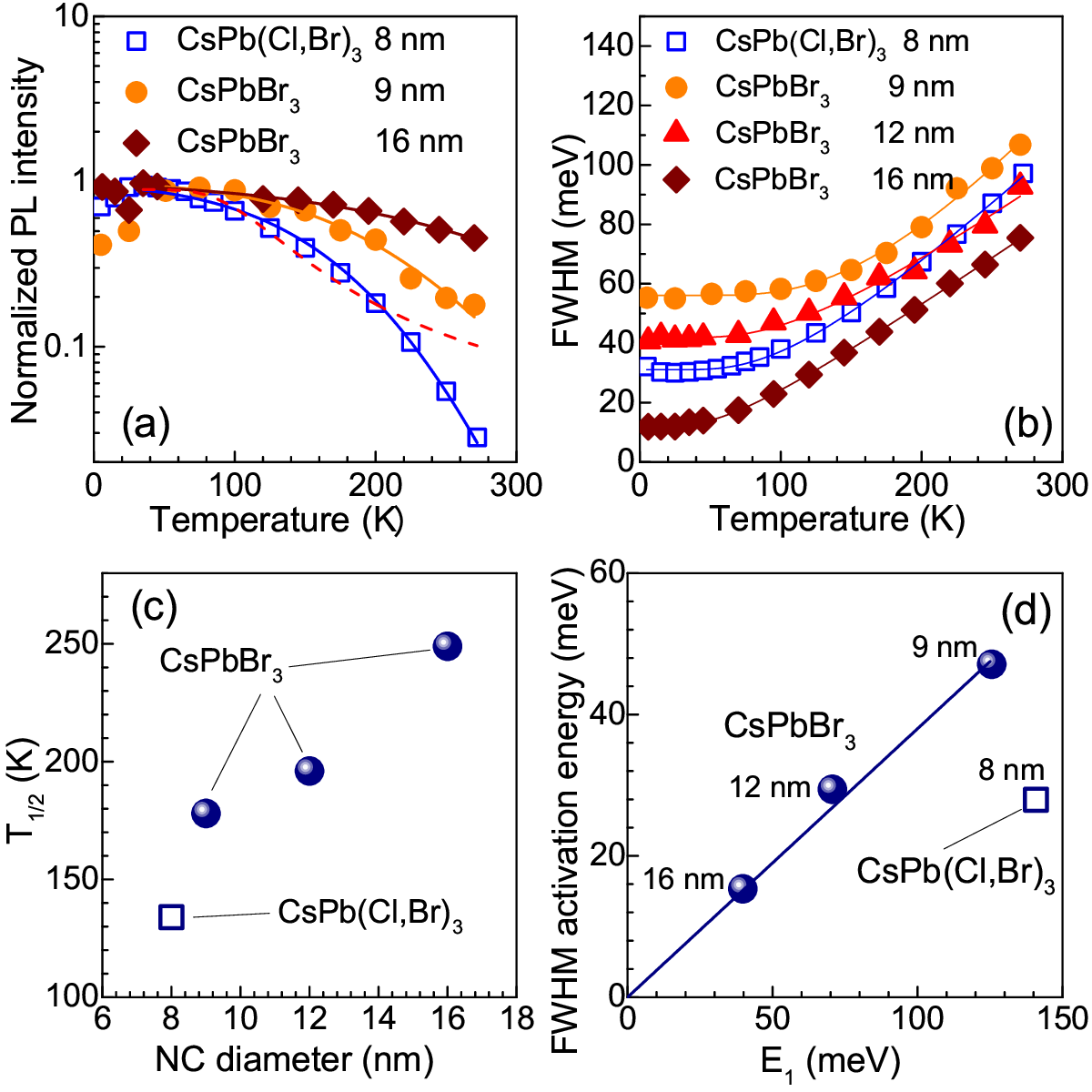}%
\caption{Temperature dependence of (a) integrated PL intensity and (b) FWHM of the PL peak for the studied samples. Solid lines show fits with Eq.~\eqref{eq:erf} and Eq.~\eqref{eq:G} to the experimentally determined intensities and FWHM, respectively. Dashed line shows a fit with Arrhenius-like equation~\eqref{eq:arr}. (c) Dependence of the temperature at which intensity decreases twice from its maximal value on the NC diameter. (d) Activation energies $E_\text{a}$ determined from the temperature dependences of FWHM as a function of NC quantum confinement energy [see Eqs.~\eqref{eq:E},\eqref{eq:G}]. Line is a linear fit. Balls in panels (c) and (d) correspond to  CsPbBr$_3$ NCs, while open square corresponds to CsPb(Cl,Br)$_3$ NCs.}
\label{fig:TDep}
\end{center}
\end{figure*}

\begin{table*}
\begin{center}
\begin{tabular}{ l  l  l  l  l  l  l  l}
    \hline
    No & Composition & Size (nm) & PL peak energy (eV) & FWHM,  $\Gamma_\text{inh}$ (meV) & $E_\text{a}$ (meV) & $\gamma_\text{ph}$ (fs$^{-1}$) & $\tau_\text{r}$ (ns) \\
    \hline
    1 & CsPb(Cl$_{0.5}$Br$_{0.5}$)$_3$ & 8 & 2.73 & 32 & 28 & 0.2 & 0.33\\
  
    2 & CsPbBr$_3$ & 9 & 2.441 & 55 & 47 & 0.5 & 1.1\\
   
    3 & CsPbBr$_3$ & 12 & 2.391 & 41 & 29 & 0.2 & 0.95\\
    
    4 & CsPbBr$_3$ & 16 & 2.361 & 12 & 15 & 0.1 & 0.55\\
    \hline
    \end{tabular}

    \caption{Summary of the NC parameters. FWHM is the PL line full width at half maximum at $T = 6$~K [equal to $\Gamma_\text{inh}$ in Eqs.~\eqref{eq:Gammainh},\eqref{eq:G}], $E_\text{a}$ is the linewidth activation energy, $\gamma_\text{ph}$ is the strength of electron--phonon coupling, and $\tau_\text{r}$ is the exciton recombination time corresponding to the fast component in the PL dynamics at $T = 6$~K.}
\label{tab}
\end{center}
\end{table*}

We show the results for three samples with CsPbBr$_3$ NCs of different sizes and for one sample with CsPb(Cl,Br)$_3$ NCs. The summary of NC parameters is
presented in table \ref{tab}. The PL spectra at a temperature of 6~K for the samples under study are shown in Fig.~\ref{fig:PLTr}(a). The energy of
the PL maximum is controlled by the band gap of the material, the quantum confinement energy of electrons and holes in NCs, and Coulomb interaction
energy between electron and hole (exciton binding energy). In particular, for the set of samples with CsPbBr$_3$ NCs different PL peak energies correspond to different NC sizes $D$. A decrease in the NC size from 16 to 9 nm for CsPbBr$_3$ NCs leads to a shift of the PL peak energy from 2.36 to 2.44 eV as a result of stronger quantum confinement. Meanwhile, the introduction of Cl further shifts the PL peak energy to 2.73 eV, as seen from the PL spectrum of CsPb(Cl,Br)$_3$ NCs with $D = 8$~nm.
It is also apparent that the full width at half maximum (FWHM) of the PL line increases with a decrease in the NC size. This is related to the fact that the low-temperature PL linewidth is determined by inhomogeneous broadening arising from the spread in the NC sizes. For small values of $D$, the PL peak energy is more sensitive to the variation of $D$, which results in larger inhomogeneous broadening, see Eqs.~\eqref{eq:E} and \eqref{eq:Gammainh} below. The linewidth is varied from 12~meV for CsPbBr$_3$ NCs with $D = 16$~nm up to 55~meV for NCs with $D = 9$~nm.

The energy structure of NCs becomes more apparent from the PL excitation spectra. Figures~\ref{fig:PLTr}(d)--\ref{fig:PLTr}(f) show the PL intensity for the samples with different NC sizes as a function of the detection energy and excitation energy. The PL has a maximum when the detection energy corresponds to the NC ground state and the excitation energy corresponds to either ground or one of the excited states. For each state, the excitation energy corresponding to the PL maximum linearly increases with the detection energy as shown by dashed lines. This is related to the distribution over NC sizes within the ensemble. Different sizes along the PL distribution corresponding to the excitation into the first excited optically active state are marked in Fig.~\ref{fig:PLTr}(d). For the sample with the largest average NC size and the smallest inhomogeneous broadening, up to five energy levels are resolved in the PL excitation spectrum [Fig.~\ref{fig:PLTr}(g)]. These levels correspond to states with different orbital quantum numbers, and their energies are well reproduced by the model of a spherical NC with infinite energy barriers presented in Modeling section. The difference between the position of the first excited optically active state determined from the PL excitation spectrum and the exciton energy of 2.32~eV in bulk CsPbBr$_3$ \cite{Belykh2019,Yakovlev2023} is used to determine the average NC size for a given sample. This method is insensitive to the Stokes shift that reduces the PL peak energy. However, we assume the same exciton binding energy for different NC sizes and quantum confined states, because its variation is small compared to the quantum confinement energy.

With temperature increase, the PL spectrum is transformed as shown in Fig.~\ref{fig:PLTr}(b) for CsPb(Cl,Br)$_3$ NCs. The energy of the PL peak shifts and the line broadens. Note that the PL spectra in Fig.~\ref{fig:PLTr}(b) are normalized to their maximal intensity at the given temperature,
while the integrated intensity varies with temperature (see below). It is noteworthy that the PL peak energy increases with temperature, in contrast to conventional semiconductors
like GaAs. At high temperatures this increase changes to a slight decrease. The transmission spectrum of the same sample experiences similar evolution 
[Fig.~\ref{fig:PLTr}(c)]: the transmission minimum, corresponding to the exciton resonance, shifts to higher energy and broadens with increasing temperature.

The temperature dependence of the integrated PL intensity for different samples is shown in Fig.~\ref{fig:TDep}(a).
Interestingly, below 40~K the intensity increases with temperature. Then it stays approximately constant in the range of 40--140~K,
and decreases at still higher temperatures. We attribute the intensity variation with temperature to carrier activation to deep-level surface traps where they can
recombine nonradiatively. This is discussed in detail in the following sections. We note that the rate of the high-temperature decrease in the PL intensity depends on the NC size and composition. We characterize this rate by the temperature $T_{1/2}$ at which the integrated PL intensity
decreases to 50\% of its maximal value. Figure~\ref{fig:TDep}(c) shows the dependence of $T_{1/2}$ on the NC diameter $D$. For the series
of CsPbBr$_3$ samples, $T_{1/2}$ increases with $D$ from 180~K for $D=9$~nm to 250~K for $D = 16$~nm. The fact that nonradiative recombination is more pronounced in smaller NCs validates the assumption about the surface origin of nonradiative centers in these samples. One can also see that the high-temperature decrease in the PL intensity in CsPb(Cl,Br)$_3$ NCs is significantly faster than in CsPbBr$_3$ NCs of similar size. This may result from the increased density of traps for NCs containing Cl.

The temperature dependence of the PL FWHM for different samples is shown in Fig.~\ref{fig:TDep}(b). The steady increase in the FWHM with $T$ is related to the
carrier activation from the ground state to excited  states via phonon absorption. The corresponding activation energies as a function of the quantum confinement energy in NCs are shown in Fig.~\ref{fig:TDep}(d).

\section{Photoluminescence dynamics}

\begin{figure*}
\includegraphics[width=1\linewidth]{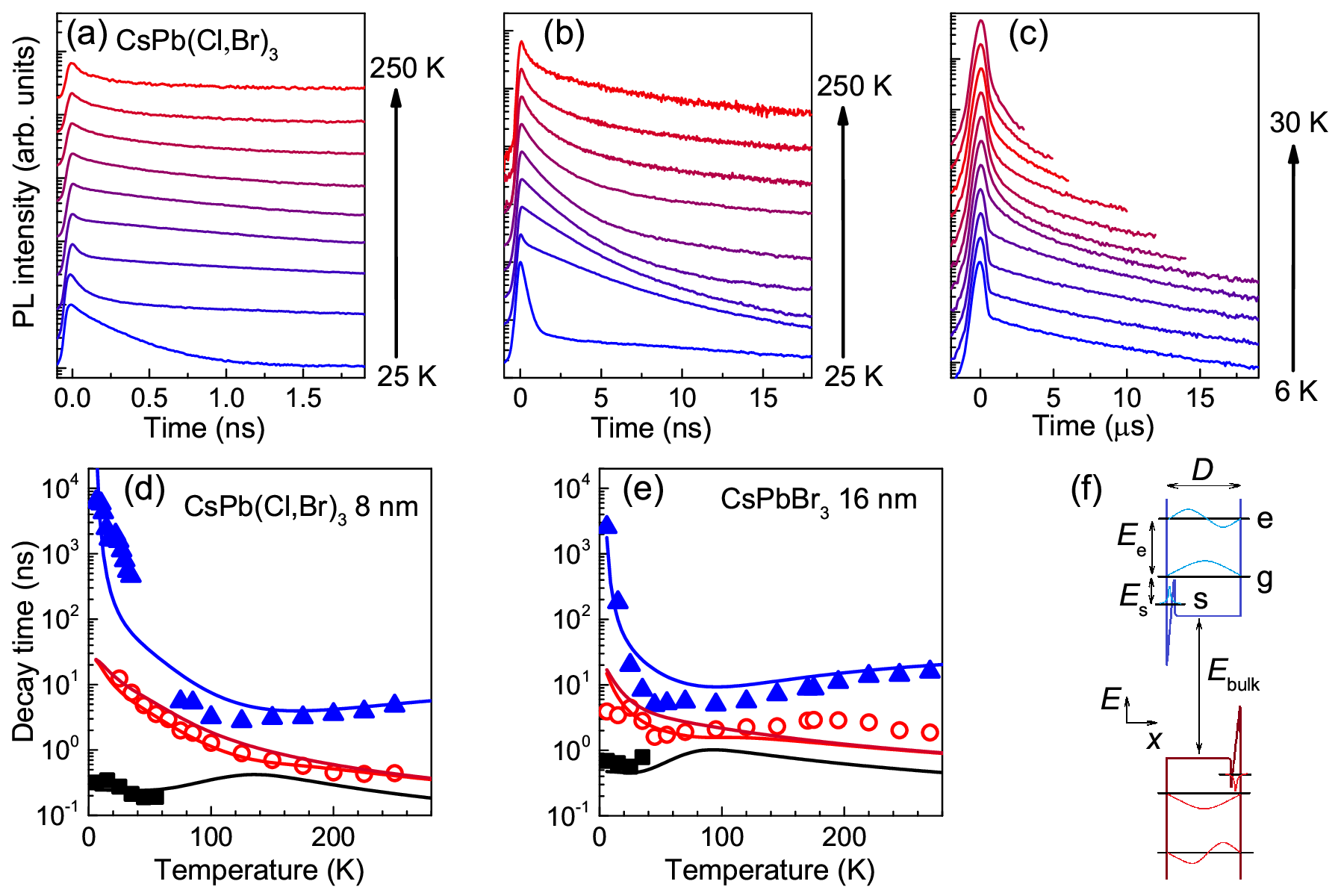}%
\caption{(a)--(c) Spectrally-integrated PL dynamics of CsPb(Cl,Br)$_3$ NCs at different temperatures in different time ranges. The curves are shifted vertically for clarity.
(d),(e) Temperature dependences  of the PL decay times for (d) 8-nm CsPb(Cl,Br)$_3$ NCs and (e) 16-nm CsPbBr$_3$ NCs. Symbols show experimental data, lines are the fits with the rate-equation model described in the Modeling section. (f) Schematic band diagram and energy levels in NCs.}
\label{fig:Dyn}
\end{figure*}

Next, we study the PL dynamics of the perovskite NCs under pulsed excitation. Figure~\ref{fig:Dyn} shows the spectrally-integrated PL dynamics of CsPb(Cl,Br)$_3$ NCs at different temperatures in different time ranges. The dynamics is characterized by several components whose amplitudes and decay times depend on temperature. The fastest component with decay time of about 0.3~ns dominates at low temperatures and disappears at higher temperatures [Fig.~\ref{fig:Dyn}(a)]. The next component is characterized by a decay time of about 10~ns at $T = 25$~K [Fig.~\ref{fig:Dyn}(b)]. It is weak at low temperatures and dominates at higher temperatures. The decay time of the slowest component is as long as 10~$\mu$s at low temperatures [Fig.~\ref{fig:Dyn}(c)] and drastically shortens with temperature increase, while its relative amplitude increases. For each temperature, a multiexponential fit was carried out to determine the characteristic decay times. The decay times of the three components as a function of temperature are shown by symbols in Figs.~\ref{fig:Dyn}(d),(e) for CsPb(Cl,Br)$_3$ and CsPbBr$_3$ NCs, respectively. It is interesting that the decay time of the slowest component first drastically decreases and then slightly increases with temperature. Qualitatively similar multicomponent PL dynamics with similar temperature dependences of the decay times is observed for other CsPbBr$_3$ NC samples.

\section{Modeling}
\subsection{Energy structure of an ideal spherical NC with infinite potential barriers}
Here we calculate the energies of the first few quantum confinement levels in a nanocrystal (NC). This calculations can be compared with resonance positions in the PL excitation spectrum [Fig.~\ref{fig:PLTr}(g)] and will be useful for understanding PL temporal dynamics behaviour and temperature dependence of the linewidth where knowledge of the separation between the ground and excited states of NC is needed. We consider the NC as a sphere with zero potential inside and infinitely high potential outside the sphere. The electron wave function (the same considerations hold for holes) can be decomposed into the product of the wave functions depending on the radial variable $R_l(r)$ and angular variables $Y_{l,m}(\theta, \phi)$, where $l$ and $m$ are the quantum numbers responsible for the square of the angular momentum and its $z$-axis component, respectively. The Schr\"{o}dinger equation for the radial wave function reads as \cite{Landau1991} 
\begin{equation}
\frac{d^2 R_l}{dr^2}+\frac{2}{r}\frac{dR_l}{dr}+ \left[k^2 - \frac{l(l+1)}{r^2}\right]R_l = 0,
\end{equation}
where $k = \sqrt{2m_\text{e} E^\text{e}/\hbar^2}$, $E^\text{e}$ is the electron energy, and $m_\text{e}$ is the effective mass.
The solution of this equation up to a normalization factor is expressed through the spherical Bessel functions: $R_l(r) = j_l(kr)$ \cite{Landau1991,Flugge2012,Elsasser1933}. The first three functions are {$j_0 (x) = \sin(x)/x$, $j_1 (x) = \sin(x)/x^2 - \cos(x)/x$, and $j_2(x) = -3\cos(x)/x^3 + (3/x^2 - 1/x) \sin(x)$}. We find $k$ and, thus, $E^\text{e}$ by imposing the boundary condition $j_l(kD/2) = 0$, where $D$ is the NC diameter. The first few states, in order of increasing energy, are 1s, 1p, 1d, 2s ..., where letters ``s'', ``p'', and ``d'' correspond to $l = 0$, 1, and 2 (the level degeneracy being $2l + 1$) and the numbers just enumerate consecutively levels with a given $l$. For the first 1s level $kD = 2\pi$ we find
\begin{equation}
E^\text{e}_1 = \frac{2\pi^2\hbar^2}{m_\text{e} D^2}.
\end{equation}
The next two levels correspond to $l = 1$ and 2 and have energies $E^\text{e}_2 \approx 2.0 E^\text{e}_1$ and $E^\text{e}_3 \approx 3.4 E^\text{e}_1$ \cite{Flugge2012}.

The quantum confinement energy for an electron--hole pair is the sum of quantum confinement energies of the electron and hole; in particular, for the ground state
\begin{equation}
E_1 = \frac{2\pi^2\hbar^2}{\mu D^2},
\label{eq:E}
\end{equation}
where $1/\mu = 1 / m_\text{e} + 1 / m_\text{h}$ is the reduced mass.  For CsPbBr$_3$ NCs we take $m_\text{e} \approx m_\text{h} \approx 0.3 m_\text{0}$ \cite{Kirstein2022}. The lowest quantum confinement energies with corresponding states and their degeneracy are given in table~\ref{tabE}.
\begin{table}
\begin{center}
\begin{tabular}{ l  l  l  l }
    \hline
    Level & Energy & States (e-h) & Dege-\\
    No & & & neracy \\
    \hline
    1 & $E_1$ & 1s-1s & 1 \\
   
    2 & $1.5 E_1$ & 1s-1p and 1p-1s & 6 \\
    
    3 & $2.0 E_1$ & 1p-1p & 9 \\
    
    4 & $2.2 E_1$ & 1s-1d and 1d-1s & 10 \\
   
    5 & $2.5 E_1$ & 1s-2s and 2s-1s & 2 \\
    
    6 & $2.7 E_1$ & 1p-1d and 1d-1p & 30 \\
    \hline
    \end{tabular}

    \caption{First few quantum confinement levels for an electron--hole pair in a spherical NC for $m_\text{e} = m_\text{h}$.}
\label{tabE}
\end{center}
\end{table}

We can compare the calculated energies with the positions of excited-state peaks in the PL excitation spectrum [Fig.~\ref{fig:PLTr}(g)]; up to five such peaks are observed for 16-nm CsPbBr$_3$ NCs. We find that the peak positions are perfectly described by the energies calculated for a spherical NC and correspond to states with the principal quantum number $n=1$ and different values of $l$. Note that the electron and hole in the NC should have the same quantum numbers to give nonzero matrix element and to contribute to the PL excitation spectrum. The states with $n=2$ have smaller $l$ (compared to the states with $n=1$ in the same energy range) and lower degeneracy, so their contribution is too weak to be visible.

We can also consider the shape of NC to be cubic, which, e.g.,  may be the case for colloidal perovskite NCs \cite{Akkerman2018}. The electron wave function in a cube is characterized by the momentum quantization numbers along the cube edges $|n_1,n_2,n_3\rangle$. In the ground state, all quantum numbers are equal to 1 for both electron and hole, and the energy is
\begin{equation}
E^\text{cube}_1 = \frac{3\pi^2\hbar^2}{2\mu L^2},
\end{equation}
where $L$ is the cube edge length. In the next energy level, all quantum numbers are still 1 except one of them, which is 2 either for the electron or hole. This gives the degeneracy of 6 for the level and energy $E^\text{cube}_2 = 1.5 E^\text{cube}_1$. The next energy level is represented by two quantum numbers out of six (three for electron and three for hole) equal to 2. This gives level degeneracy of 15 and energy $E^\text{cube}_3 = 2.0 E^\text{cube}_1$. We see that the energy-level structure of the lowest states is not very sensitive to the NC shape. In particular, the separations between the first three levels, measured in units of $E_1$, are the same for spherical and cubic NCs.
Note that, in the following model of electron--hole dynamics, we consider the first three energy levels only.

\subsection{PL intensity quenching with temperature}
To understand the observed decrease of NCs PL with temperature, first of all we note the dominant role of the NC surface, as the decrease rate becomes higher for smaller NCs.  
Here we assume that the NC becomes nonradiative due to deep traps present at the NC surface with density $\sigma$. A carrier can be trapped once it overcomes a potential barrier with height $E_\text{tr}$, and then it recombines nonradiatively with the other carrier or mediates Auger recombination. Then we take into account that the rate of nonradiative recombination in an inhomogeneous NC ensemble may vary from one NC to another. Thus, the rate of carrier trapping and nonradiative recombination is $\gamma_\text{nr}\exp(-E_\text{tr}/k_\text{B}T)$, and a NC is ``dark'' if $\gamma_\text{nr}\exp(-E_\text{tr}/k_\text{B}T) \gg \gamma_\text{r}$ and is
``bright'' if $\gamma_\text{nr}\exp(-E_\text{tr}/k_\text{B}T) \ll \gamma_\text{r}$; here, $\gamma_\text{r}$ and $\gamma_\text{nr}$ are the radiative and the high-temperature nonradiative recombination rates, respectively. The intermediate regime $\gamma_\text{nr}\exp(-E_\text{tr}/k_\text{B}T) \sim \gamma_\text{r}$ takes place only in a narrow temperature range, provided that $\gamma_\text{nr} \gg \gamma_\text{r}$. The smooth decrease in the PL intensity with temperature occurs because of the spread in the trap potential barrier heights $E_\text{tr}$ \cite{Savchenko2022}. So we can assume that the given NC is ``switched off'' in PL when the temperature reaches the value $T = E_\text{tr} /k_\text{B} \ln(\gamma_\text{nr}/\gamma_\text{r})$.

Now, let us cast these considerations in a more quantitative form. Assume that the NCs are spherical and the number of traps $n$ in an arbitrary NC is described by a Poissonian distribution with mean value of $\pi D^2 \sigma$:
\begin{equation}
p_n = \frac{(\pi D^2 \sigma)^n}{n!}\exp(-\pi D^2 \sigma).
\end{equation}
Let the rate of charge-carrier capture by a single trap be $\gamma_\text{tr}\exp(-E_\text{tr}/k_\text{B}T)$ and the trap barrier energies be distributed according to the Gaussian function with mean $\bar E_\text{tr}$ and variance $\Gamma_\text{tr}^2$: 
\begin{equation}
f(E_\text{tr}) = \frac{1}{\sqrt{2\pi}\Gamma_\text{tr}}\exp\left(-\frac{[E_\text{tr} - \bar E_\text{tr}]^2}{2\Gamma_\text{tr}^2}\right).
\end{equation}
We assume $\Gamma_\text{tr} \ll \bar E_\text{tr}$, so that $\int_0^\infty f(E_\text{tr}) dE_\text{tr} \approx \int_{-\infty}^\infty f(E_\text{tr}) dE_\text{tr} = 1$.
The probability that a charge carrier in a NC containing one trap will recombine radiatively at temperature $T$, i.e. that the trap activation energy is higher than $k_\text{B}T\ln(\gamma_\text{tr}/\gamma_\text{r})$, is
\begin{multline}
F(T) = \int_{k_\text{B}T\ln(\gamma_\text{tr}/\gamma_\text{r})}^{\infty} f(E_\text{tr}) dE_\text{tr} \approx \\
\frac{1}{2} - \frac{1}{2} \text{erf}\left(\frac{k_\text{B}T\ln(\gamma_\text{tr}/\gamma_\text{r}) - \bar E_\text{tr}}{\sqrt{2}\Gamma_\text{tr}}\right),
\end{multline}
where erf is the error function:
\begin{equation}
\text{erf}(x) = \frac{2}{\sqrt{\pi}}\int_0^x \exp(-t^2)dt,
\end{equation}
$\text{erf}(-\infty) = -1$, $\text{erf}(\infty) = 1$.
The cutoff temperature $\tilde{T}$ above which the capture of charge carriers at nonradiative recombination centers becomes faster than radiative recombination in a NC containing $n$ such centers decreases logarithmically with $n$ as
$\tilde{T} = \bar{E_\text{tr}}/{k_\text{B}\ln(n \gamma_\text{tr}/\gamma_\text{r}})$.
However, if $n \ll \gamma_\text{tr}/\gamma_\text{r}$, we can disregard this dependence and take $\tilde{T} = \bar{E_\text{tr}}/{k_\text{B}\ln(\gamma_\text{tr}/\gamma_\text{r}})$.
Then, the probability $F_n(T)$ that a NC containing $n$ traps is still bright at temperature $T$ is
\begin{equation}
F_n(T) \approx F(T)^n \approx \\
\left[\frac{1}{2} - \frac{1}{2} \text{erf}\left(\frac{k_\text{B}T\ln(\gamma_\text{tr}/\gamma_\text{r}) - \bar E_\text{tr}}{\sqrt{2}\Gamma_\text{tr}}\right)\right]^n.
\end{equation}
The PL intensity is proportional to the probability $P(T)$ that an arbitrary NC is still bright at temperature $T$: 
\begin{multline}
I(T) \propto P(T) = \sum_{n=0}^\infty F_n(T) p_n \approx \\
\exp\left(-\frac{\pi D^2 \sigma}{2}\left\lbrace 1 + \text{erf}\left[ \frac{k_\text{B}T\ln(\gamma_\text{tr}/\gamma_\text{r}) - \bar E_\text{tr}}{\sqrt{2}\Gamma_\text{tr}} \right]\right\rbrace\right).
\label{eq:erf}
\end{multline}

It is important to compare the dependences $I(T)$ for NCs with different sizes. We assume that $\bar E_\text{tr}$ and $\Gamma_\text{tr}$ are independent of $D$, while for $\gamma_\text{r}$ we can use the experimental values neglecting its temperature dependence (note that $\gamma_\text{r}$ is under logarithm). To determine the dependence of $\gamma_\text{tr}$ on $D$ we note that $\gamma_\text{tr}$ is proportional to the probability of finding the electron (or the hole) near a trap, in the vicinity of the surface characterized by some small distance $\delta r \ll D$: 
\begin{equation}
\gamma_\text{tr} \propto R^2_0(D/2 - \delta r),
\end{equation}
where $R_0 =  \sin(2\pi r/D) / \sqrt{\pi D} r$ is the normalized radial wave function for $l=0$. Therefore,
$\gamma_\text{tr} \propto 4 \pi \delta r^2 / D^5$. The same dependence on $D$ is obtained if we assume a finite height of the NC potential barriers and calculate the squared wave function exactly at the NC surface. Thus,
\begin{equation}
\gamma_\text{tr} = \eta D^{-5},
\label{eq:SizeDep}
\end{equation}
where $\eta$ is a constant factor. The mean total nonradiative recombination rate in NCs is
\begin{equation}
\gamma_\text{nr} ={\pi D^2 \sigma}\gamma_\text{tr} \propto D^{-3}.
\end{equation}

We fit the PL intensity temperature dependences at $T > 50$~K for the three CsPbBr$_3$ samples [Fig.~\ref{fig:TDep}(a)] with Eqs.~\eqref{eq:erf} and \eqref{eq:SizeDep} and determine the following parameters common for the samples: $\sigma = 0.024$~nm$^{-2}$, $\eta = 3\times 10^4$~ps$^{-1}$nm$^5$, $\bar E_\text{tr} = 180$~meV and $\Gamma_\text{tr} = 70$~meV. The fit shows good agreement with the experimental data. Using these values, we find that the average total number of traps $\pi D^2 \sigma$ is 6, 11, and 19, while the (high-temperature) capture time $\gamma_\text{tr}^{-1} = 2$, 8, and 35~ps for NCs with $D = 9$, 12, and 16~nm, respectively. The same parameters except for $\sigma = 0.077$~nm$^{-2}$ ($\pi D^2 \sigma$ = 15) and $\gamma_\text{tr}^{-1} = 1$ ps are used to fit the intensity temperature dependence for CsPb(Cl,Br)$_3$ sample [Fig.~\ref{fig:TDep}(a)]. Thus, the introduction of Cl to the NCs leads to an increase in the density of surface traps as well as the capture rate per trap and, therefore, faster temperature quenching of the PL intensity.

We note that the condition $n \ll \gamma_\text{tr}/\gamma_\text{r}$ is not satisfied for the sample with the largest NCs. This points to the fact that, owing to a number of approximations made, the above analysis is not quantitatively exact. However, it demonstrates the self-consistency of the general picture, and the good agreement between the experimental data and the calculated curves underscores the adequacy of the model.

For comparison, we also show the fit with Arrhenius-like equation [red dashed line in Fig.~\ref{fig:TDep}(a)]
\begin{equation}
I(T) = \frac{I(0)}{1 + \kappa \exp(-E_\text{a}/k_\text{B}T)},
\label{eq:arr}
\end{equation} 
where $\kappa$ in the activation rate and $E_\text{a}$ is the activation energy which equals 48 meV in this case. This equation is often used in the literature to fit the temperature dependence of the PL intensity \cite{Ai2017}.
This fit shows worse agreement with the experimental data than Eq.~\eqref{eq:erf}, especially at high temperatures.

\subsection{Model for electron--hole recombination dynamics in NCs}
To explain the complex PL dynamics of the NCs, we take into account the following considerations. In the low-temperature regime, the excited states do not contribute to the PL dynamics since they are separated by several tens of meV from the ground state. Nevertheless, even at low temperatures the dynamics is rather complicated, which cannot be explained by recombination from the ground state only. The presence of a long component with a lifetime strongly dependent on $T$ [Fig.~\ref{fig:Dyn}(d),(e)] suggests the existence of dark exciton states. A long-lived component demonstrating similar behaviour with temperature was found for CsPbBr$_3$ and CsPbI$_3$ NCs and was attributed to the activation from a dark exciton state lying several meV below the bright state \cite{Rossi2020}. The dark state may be attributed to the spin-forbidden exciton state \cite{Shornikova2018}. However, it cannot explain the existence of the intermediate component with a decay time of about 10~ns. A dark exciton state was indeed observed in the study of charge-carrier spin dynamics in the same CsPb(Cl,Br)$_3$ sample as presently investigated \cite{Belykh2022}. Based on an extremely small value of the electron--hole exchange splitting and the fact that the PL lifetime was independent of the magnetic field, this state was attributed to the spatially indirect exciton which forms when one of the carriers is captured by a potential trap at the NC surface. Building upon this interpretation, we assume that an electron and a hole shallow-level surface trap states \textbf{s} that are separated by energy $E_\text{s}$ from the respective particle's ground state \textbf{g} [Fig.~\ref{fig:Dyn}(f)] exist in the NC. We also consider the first excited, \textbf{e}, level for each carrier [Fig.~\ref{fig:Dyn}(f)], and neglect the contribution from the higher levels. The separation between \textbf{g} and \textbf{e} levels $E_\text{ge}$ is few tens of meV, so contributions from excited states are expected only at high temperatures. Taking into account that the effective masses of electrons and holes in CsPbBr$_3$ perovskites are almost the same (and equal to $0.3 m_\text{0}$) \cite{Kirstein2022}, we assume that the separations between the levels in the energy spectra of electrons and holes are the same, too. So $E_\text{s}$ stands for the separation between the \textbf{s} and \textbf{g} levels and $E_\text{e}$ stands for the separation between \textbf{g} and \textbf{e} levels for both electrons and holes. 

We can write the following rate equations for the exciton populations $n_{\alpha,\beta}$, corresponding to the electron at a level $\alpha = s, g, e$ and hole at a level $\beta = s, g, e$:
\begin{align}
\frac{d n_{\text{s},\beta}}{dt} = \gamma_\text{sg} (1+N_\text{sg}) n_{\text{g},\beta} + \gamma_\text{se} (1+N_\text{se}) n_{\text{e},\beta} \nonumber\\
- \gamma_\text{sg} N_\text{sg} n_{\text{s},\beta} - G \gamma_\text{se} N_\text{se} n_{\text{s},\beta},     \\
\frac{d n_{\text{g},\beta}}{dt} = \gamma_\text{sg} N_\text{sg} n_{\text{s},\beta} + \gamma_\text{ge} (1+N_\text{ge}) n_{\text{e},\beta} \nonumber\\
- \gamma_\text{sg} (1+N_\text{sg}) n_{\text{g},\beta} - G \gamma_\text{ge} N_\text{ge} n_{\text{g},\beta} - \gamma_\text{r} n_{\text{g},\beta} \delta_{g,\beta},\\
\frac{d n_{\text{e},\beta}}{dt} = G \gamma_\text{se} N_\text{se} n_{\text{s},\beta} + G \gamma_\text{ge} N_\text{ge} n_{\text{g},\beta} \nonumber\\
- \gamma_\text{se} (1+N_\text{sg}) n_{\text{e},\beta} - \gamma_\text{ge} (1+N_\text{ge}) n_{\text{e},\beta}
\end{align}
and similar equations for exchanged indices in $n_{\alpha,\beta}$.
Here, $\delta_{g,\beta} = 1$ for $\beta = g$ and $\delta_{g,\beta} = 0$ for $\beta \neq g$; $N_\text{sg} = [\exp(E_\text{s}/k_\text{B}T)-1]^{-1}$, $N_\text{ge} = [\exp(E_\text{e}/k_\text{B}T)-1]^{-1}$, and $N_\text{se} = [\exp(\{E_\text{s}+E_\text{e}\}/k_\text{B}T)-1]^{-1}$ are populations of phonons at energies corresponding to the difference between the respective energy levels; $\gamma_\text{sg}$, $\gamma_\text{ge}$, and $\gamma_\text{se}$ are the phonon-related relaxation rates between levels \textbf{s}--\textbf{g}, \textbf{g}--\textbf{e}, and \textbf{s}--\textbf{e} (we assume them equal for electrons and holes); and $\gamma_\text{r}$ is the radiative recombination rate in the ground state. 
The factor $G$ represents the effective number of excited states. In the simplest case, $G$ is the degeneracy of the excited state (3 in our case). However, in the fit we take $G = 8$, which in part accounts for the large number of excited states which are not included explicitly in the model. For simplicity, we assume only one shallow trap per NC for electrons and one for holes. Introduction of multiple traps would introduce more parameters to the model, while not leading to any qualitative changes of the results. We assume that radiative recombination only takes place when both the electron and hole are in the \textbf{g} state. Indeed, the overlap of wavefunctions for \textbf{s}--\textbf{s}, \textbf{s}--\textbf{g}, \textbf{e}--\textbf{g}, and \textbf{s}--\textbf{e} states is vanishingly small, while the occupancy of the \textbf{e}--\textbf{e} state (1p--1p) is low (being the product of the electron and hole excited-state occupancies) and, taking into account selection rules, recombination from this state is further reduced by the degeneracy factor $2l+1$. To fit the experimental data in Fig.~\ref{fig:Dyn}(d) for CsPb(Cl,Br)$_3$, we use the following parameters: $\gamma_\text{r} = 4$~ns$^{-1}$, $\gamma_\text{sg} = 0.04$~ns$^{-1}$, $\gamma_\text{se} = 0.5$~ns$^{-1}$, and $E_\text{s} = 2.2$~meV. For the excited state energy separation $E_\text{e}$ and the transition rate $\gamma_\text{ge}$, we use the values determined from the fit of the temperature dependence of FWHM with Eq.~\eqref{eq:G} (see below): $E_\text{e} = 28$~meV and $\gamma_\text{ge} = \gamma_\text{ph} / G = 30$~ps$^{-1}$ for CsPb(Cl,Br)$_3$.

The model gives four groups of relaxation times. In the first group, the relaxation times are of the order of $\gamma_\text{ge}^{-1} = 30$~fs and are not revealed in our PL dynamics experiment. The other three groups follow the branches of the experimentally measured decay times. According to the model, at low temperatures the shortest decay time (of these three) is close to the recombination time $\gamma_\text{r}^{-1} + \gamma_\text{sg}^{-1} \approx \gamma_\text{r}^{-1} = 0.25$~ns. The intermediate time is close to the relaxation time from the ground state to the surface trap $\gamma_\text{sg}^{-1} = 25$~ns. The longest time, which is strongly dependent on temperature, describes the activation of surface-trapped carriers to the ground state and can be estimated as $(\gamma_\text{sg} N_\text{sg})^{-1}$.

To fit the temperature dependence of the PL decay times for the largest 16-nm CsPbBr$_3$ NCs [Fig.~\ref{fig:Dyn}(e)], we also used $E_\text{e}$ and $\gamma_\text{ge}$ determined from the FWHM temperature dependence ($E_\text{e} = 15$~meV and $\gamma_\text{ge} = \gamma_\text{ph} / G = 10$~ps$^{-1}$), while the other parameters are $\gamma_\text{r} = 2$~ns$^{-1}$, $\gamma_\text{sg} = 0.05$~ns$^{-1}$, $\gamma_\text{se} = 0$~ns$^{-1}$, and  $E_\text{s} = 1.3$~meV. We note the almost twice smaller depth of the shallow surface traps $E_\text{s}$ for the more homogeneous sample. The main fit parameters are summarized in table~\ref{tabDyn}. 
\begin{table}
\begin{center}
\begin{tabular}{ l  l  l  l}
    \hline
    Sample & $\gamma_\text{r}$ & $\gamma_\text{sg}$ & $E_\text{s}$ \\
      & ns$^{-1}$ & ns$^{-1}$ & meV \\
    \hline
    CsPb(Cl,Br)$_3$ (8 nm) & 4 & 0.04 & 2.2 \\
    
    CsPbBr$_3$ (16 nm) & 2 & 0.05 & 1.3 \\
    \hline
    \end{tabular}

    \caption{Main parameters used to fit the temperature dependence of the PL decay times.}
\label{tabDyn}
\end{center}
\end{table}

\section{Discussion}
Here we outline a systematic picture of the PL properties of perovskite NCs, i.e., of the broadening and thermal quenching of the PL lines and the PL dynamics.
The most straightforward and expected observation is the increase in the PL peak energy and inhomogeneous broadening of the PL line with a decrease in the average size of NCs [Fig.~\ref{fig:PLTr}(a)].
The dependence of the PL energy on the NC size can be evaluated from the model of a spherically symmetric quantum dot with abrupt infinite potential barriers presented above [see Eq.~\eqref{eq:E}]. Here we neglect the excitonic effect. Note that, in this model, the separation between the ground and the first excited optically active states is equal to the quantum confinement energy of the ground state $E_1$, and this fact is used to determine the NC average size from the PL excitation spectrum.

Then, the inhomogeneous broadening, which determines the PL linewidth $\Gamma_\text{inh}$ at low temperatures and arises from the spread of the NC sizes $\Delta D$, can be evaluated using Eq.~\eqref{eq:E} as $\Gamma_\text{inh} = |\partial E_1/\partial D|\Delta D$:
\begin{equation}
\Gamma_\text{inh} = \frac{4\pi^2\hbar^2 \Delta D}{\mu D^3}.
\label{eq:Gammainh}
\end{equation}
So, the narrowest PL line is obviously obtained in the largest NCs, emitting at the lowest energies; for 16-nm CsPbBr$_3$ NCs,  the FWHM is as small as 12~meV.
It is interesting that inhomogeneous broadening of the PL line in CsPb(Cl,Br)$_3$ NCs is twice smaller than that in CsPbBr$_3$ NCs of similar size.

The PL linewidth increases with temperature [Fig.~\ref{fig:TDep}(b)]. This is explained by the dephasing of the exciton ground state in the NCs caused by the phonon-assisted transitions of electrons and holes to higher energy levels. This behavior is commonly described by the relation
\begin{equation}
\Gamma (T) = \Gamma_\text{inh} + \gamma_\text{ph}\frac{1}{\exp(E_\text{a}/k_\text{B}T)-1}.
\label{eq:G}
\end{equation}
Here, $\gamma_\text{ph}$ characterizes the strength of electron--phonon coupling, the energy $E_\text{a}$ that determines the Bose--Einstein occupancy factor $\left(\exp(E_\text{a}/k_\text{B}T)-1\right)^{-1}$ is the energy of the polar longitudinal-optical (LO) phonon mode, and we neglect the contribution from acoustical phonons. The values obtained from the fit are given in table~\ref{tab}. This relation is well established for systems with continuous energy spectrum, where scattering by LO phonons simply limits the lifetime of radiative excitonic states (see, e.g., \cite{Rudin1990}). This relation also describes systems with discrete energy spectrum, such as epitaxial and colloidal quantum dots \cite{Bayer2002,Valerini2005}. Although no real transitions can generally be induced by LO phonons in this case, dephasing caused by virtual transitions involving excited states (i.e., by the off-diagonal terms of the electron--phonon coupling Hamiltonian) leads to the behavior numerically well described by Eq.~\eqref{eq:G} \cite{Muljarov2007}.

The perovskite crystal lattice features a great number of optical phonon modes, and it is not {\it a priori} clear  if this simple formula can be valid in this case. However, a number of studies indicate that only one of the optical-phonon modes of this lattice possesses a large dipole moment and, therefore, is dominant in the Fr\"{o}hlich carrier--phonon interaction \cite{Perez-Osorio2015,Wright2016,Iaru2021}. For CsPbBr$_3$, the energy of this mode is about $E_\text{LO}^\text{eff}$ = 20 meV \cite{Iaru2021}. Correspondingly, it was found that the temperature dependence of the PL linewidth in bulk and NC perovskite materials can be described by Eq.~\eqref{eq:G} with $E_\text{a}$ on the same order of magnitude (e.g. \cite{Wright2016,Ramade2018}).

The PL linewidth in our samples [Fig.~\ref{fig:TDep}(b)] can also be well fitted with Eq.~\eqref{eq:G}. The values obtained from the fit are given in table~\ref{tab}. However, one can see that the activation energy $E_\text{a}$ is not simply a constant that can be associated with the energy of a specific phonon mode. Only for the largest CsPbBr$_3$ NCs $E_\text{a}$ is close to $E_\text{LO}^\text{eff}$, while it is noticeably larger for smaller NCs.
The dependence of $E_\text{a}$ on the electron--hole pair quantum-confinement energy $E_1$ in the NCs [the latter is calculated by Eq.~\eqref{eq:E}] is shown in Fig.~\ref{fig:TDep}(d) and, for the series of CsPbBr$_3$ samples, follows a linear relation $E_\text{a} = 0.4 E_1$. We note that, in a quantum dot with infinite potential barriers, the carrier ground-state energy with respect to the bottom of the potential is almost equal to the separation between the ground and first excited states (see Modeling section). Then, for an electron--hole pair, the energy of the first excited state (when either of carries is on the next energy level, this state is not optically active) is equal to $1.5 E_1$. Therefore, the value of $E_\text{a}$ is close to the separation between the ground and first excited states of an electron--hole pair, which equals $0.5 E_1$.
The significance and, to that end, the universality of this relation (note that the point corresponding to the CsPb(Cl,Br)$_3$ sample deviates considerably) is the subject of further investigation. However, it may be speculated that this behavior occurs because there actually exist contributions from more than just one phonon mode to the dephasing rate, and the mode resonant with the energy of transitions from the ground to the first excited state has the greatest impact on $\Gamma (T)$. Hence, this transition energy enters Eq.~\eqref{eq:G}.

The decrease in the PL intensity with increasing temperature [Fig.~\ref{fig:TDep}(a)] is a general feature that is observed in many semiconductor structures, including perovskite NCs \cite{Ai2017,Shi2019,Wu2019,Lohar2018,Ye2019,Pham2022,Wei2016} and single crystals \cite{Wolf2018}. It is often attributed to the activation of nonradiative recombination with temperature. However, the latter process is accompanied by a corresponding shortening of the PL decay time. This is not the case in our experiments. 
For example, for CsPb(Cl,Br)$_3$ NCs the PL intensity drops by more than an order of magnitude as the temperature is increased from 100 to 270~K [Fig.~\ref{fig:TDep}(a)]. Meanwhile, the short decay time decreases by only a factor of 3, and the long decay time even increases [Fig.~\ref{fig:Dyn}(d)]. To understand the observed quenching we consider an inhomogeneous NCs ensemble with a large spread in the rates of nonradiative recombination. The PL dynamics in ``bright'' NCs with the slowest nonradiative recombination will still be determined by radiative processes, while PL in ``dark'' NCs with the fastest nonradiative recombination will decay so rapidly that we cannot detect it. Under reasonable assumptions, discussed in the Modeling section, most of the NCs will fall into one of two categories with the boundary between them shifting with temperature. Therefore, an increase in temperature will just lead to an increase in the number of NCs that recombine predominantly nonradiatively and do not contribute to PL and, thus, have no effect on the PL dynamics. This may take place in the following two scenarios which imply the existence of deep-level surface traps with relatively high potential barriers. As the temperature is increased, one of the carriers may be captured by such a trap. In the first scenario, the carrier captured by a surface trap recombines nonradiatively with the carrier remaining in the core. In the second scenario, the NCs contain resident carriers that are captured by the surface at higher temperatures and assist the Auger nonradiative recombination of photoexcited electron--hole pairs, which is very efficient for NCs with sharp confining potential \cite{Cragg2010}. Similar explanation was proposed for the temperature quenching of the quantum yield in CdSe/CdS colloidal NCs in ref.~\cite{Javaux2013}, where the PL decay time showed an increase with temperature. 

The reduction of the PL intensity with temperature is fitted with the model presented above [solid lines in Fig.~\ref{fig:TDep}(a)] assuming a Gaussian distribution of trap barrier energies. In this model nonradiative recombination rate $\gamma_\text{nr}$ decreases as the 3rd power of the NC diameter.  The fit shows good agreement with experimental data, better than Arrhenius-like equation \eqref{eq:arr} which is often used in the literature to fit temperature dependences of the PL intensity [red dashed line in Fig.~\ref{fig:TDep}(a)] \cite{Ai2017,Wu2019}.

To explain the multiscale PL dynamics of the NCs, we take into account shallow trap states besides ground and excited states for each type of carriers.
Thus, at low temperatures, the following electron–hole pair states should be included in the analysis: (i) both the electron and hole in the ground state of the NC, (ii) they both in the respective trap states, and (iii) either carrier is in the ground and the other in the trap state. The rate-equation model of the preceding section, considering phonon-induced transitions between these states, yields reasonably good fits for the temperature dependences of the experimentally observed decay times, shown by lines in Fig.~\ref{fig:Dyn}(d),(e). As far as the electron--hole wavefunction overlap is small in the 2nd and 3rd cases, we assume that only state (i) is radiative, and the shortest PL component, according to the model, corresponds to radiative recombination and to relaxation from this state to state (iii) with the rate $\gamma_\text{r}+\gamma_\text{sg} \approx \gamma_\text{r}$. The intermediate PL component corresponds to the relaxation of either carrier from the NC ground to the surface trap state with the rate $\gamma_\text{sg}$, while the second carrier is already in the trap. Note that model predicts two PL components with very close decay times in the intermediate range [red curves in Fig.~\ref{fig:Dyn}(d),(e)] which can hardly be separated in the experiment. Important that the model fails to predict the intermediate component if a trap for only one type of carriers is present. Finally, the long component corresponds to carrier activation from the trap to the NC ground state with time $\gamma_\text{sg}^{-1}\exp(E_\text{s}/T)$ (which leads to recombination whenever both carriers happen to be activated simultaneously). We also mention that the initial increase in the integrated PL intensity with $T$ [Fig.~\ref{fig:TDep}(a)] may be related to the nonradiative recombination of carriers in these shallow traps. Indeed, at low temperatures carriers spend most of their time in trap states (having more chances to recombine nonradiatively) and are activated to the NC ground state at higher temperatures.

Another pronounced feature of the PL dynamics is the increase in the decay time at high temperatures. This is a common feature for systems with continuous density of states such as quantum wells \cite{Feldmann1987,Belykh2015}. It is related to the filling with temperature of the reservoir of nonradiative exciton states with momenta beyond the light cone, while only excitons with momenta within the light cone and, thus, with low energy can recombine radiatively. We note that an increase in the PL lifetime with temperature was also observed in nanocrystals \cite{Javaux2013}. Much like in other systems, we attribute this increase to the activation of carriers to excited states, where their recombination is inhibited. As the ground-state occupancy decreases with temperature, so does the recombination rate.

\section{Conclusions}
To conclude, we have studied the optical properties of CsPbBr$_3$ and CsPb(Cl,Br)$_3$ NCs in a glass matrix: stationary PL and PL excitation spectra, transmission spectra, and PL dynamics. We have shown that an increase in the NC size results in weaker inhomogeneous broadening of the PL line corresponding to excitons. The exciton linewidth has activation-like dependence on temperature with an activation energy close to the interlevel separation in the NCs. We have observed PL quenching with temperature that is not accompanied by an increase in the PL decay rate that should be expected for the activation of nonradiative recombination. We show that this quenching is related to the existence of deep surface traps and becomes more pronounced with a decrease in the NC size and the introduction of Cl. For the largest CsPbBr$_3$ NCs, the PL intensity decreases by only 50\% upon an increase in temperature from 6 to 270 K, which gives evidence of their high quality. The fact that considerable thermal quenching of PL occurs only in smaller NCs indicates that the number of nonradiative recombination centers in the bulk is small and they are mostly located at the surface. The PL dynamics of the studied NCs at low temperatures is characterized by three time scales  on the order of 1 ns, 10 ns, and 1 ${\mu}$s, respectively. This dynamics is described by a model considering relaxation of both electrons and holes to shallow traps and their activation from the ground to the excited states.

\section{Acknowledgements}
We acknowledge financial support by the Ministry of Science and Higher Education of the Russian Federation, Contract No. 075-15-2021-598 at the P.N.~Lebedev Physical Institute.
The work of M.S.K. (sample characterization) was supported by the Saint Petersburg State University through Research Grant No.~94030557.

\end{document}